\documentclass[twocolumn]{emulateapj}

\usepackage{float}

\usepackage[caption=false]{subfig}

\begin{document}

\title{The Main Sequence at $\lowercase{z} \sim 1.3$ contains a sizable fraction of galaxies with compact star formation sizes: a new population of early post-starbursts?} 

\author{A. Puglisi \altaffilmark{1,2}} 
\author{E. Daddi \altaffilmark{1}} 
\author{D. Liu \altaffilmark{3}} 
\author{F. Bournaud \altaffilmark{1}}
\author{J. D. Silverman \altaffilmark{5}} 
\author{C. Circosta \altaffilmark{4}} 
\author{A. Calabr\`o \altaffilmark{1}}
\author{M. Aravena \altaffilmark{19}} 
\author{A. Cibinel \altaffilmark{16}} 
\author{H. Dannerbauer \altaffilmark{14, 15}} 
\author{I. Delvecchio \altaffilmark{1}}
\author{D. Elbaz \altaffilmark{1}} 
\author{Y. Gao \altaffilmark{13}}
\author{R. Gobat \altaffilmark{17}}
\author{S. Jin \altaffilmark{19}}
\author{E. Le Floc'h \altaffilmark{1}}
\author{G. E. Magdis \altaffilmark{10, 11, 12}}
\author{C. Mancini \altaffilmark{6, 2}}
\author{D. A. Riechers \altaffilmark{7,8}}
\author{G. Rodighiero \altaffilmark{6,2}}
\author{M. Sargent \altaffilmark{16}} 
\author{F. Valentino \altaffilmark{10}} 
\author{L. Zanisi \altaffilmark{9}} 

\affiliation{
\\1, CEA, IRFU, DAp, AIM, Universit\'e Paris-Saclay, Universit\'e Paris Diderot, Sorbonne Paris Cit\'e, CNRS, F-91191 Gif-sur-Yvette, France
\\2, INAF-Osservatorio Astronomico di Padova, Vicolo dell'Osservatorio, 5, 35122 Padova, Italy
\\3, Max Planck Institute for Astronomy, Konigstuhl 17, D-69117 Heidelberg, Germany
\\4, ESO, Karl-Schwarschild-Stra{\ss}e 2, 85748 Garching bei M{\"u}nchen, Germany
\\5, Kavli Institute for the Physics and Mathematics of the Universe (WPI), Todai Institutes for Advanced Study, the University of Tokyo, Kashiwanoha, Kashiwa, 277-8583, Japan
\\6, Dipartimento di Fisica e Astronomia, Universit{\`a} di Padova, vicolo dell’Osservatorio 2, 35122 Padova, Italy
\\7, Cornell University, Space Sciences Building, Ithaca, NY 14853, USA
\\8, Max-Planck-Institut f\"ur Astronomie, K\"onigstuhl 17, D-69117 Heidelberg, Germany
\\9, Department of Physics and Astronomy, University of Southampton, Highfield SO17 1BJ, UK
\\10, Cosmic Dawn Center at the Niels Bohr Institute, University of Copenhagen and DTU-Space, Technical University of Denmark
\\11, Niels Bohr Institute, University of Copenhagen, DK-2100 Copenhagen $\O$
\\12, Institute for Astronomy, Astrophysics, Space Applications and Remote Sensing, National Observatory of Athens, 15236, Athens, Greece
\\13, Purple Mountain Observatory/Key Lab of Radio Astronomy, Chinese Academy of Sciences, Nanjing 210034, PR China
\\14, Instituto de Astrofisica de Canarias (IAC), E-38205 La Laguna, Tenerife, Spain 
\\15, Universidad de La Laguna, Dpto. Astrofisica, E-38206 La Laguna, Tenerife, Spain
\\16, Astronomy Centre, Department of Physics and Astronomy, University of Sussex, Brighton, BN1 9QH, UK
\\17, Instituto de Fisica, Pontificia Universidad Cat{\'o}lica de Valpara{\'i}so, Casilla 4059, Valparaiso, Chile
\\18, Key Laboratory of Modern Astronomy and Astrophysics in Ministry of Education, School of Astronomy and Space Science, Nanjing University, Nanjing 210093, China
\\19, N{\'u}cleo de Astronom{\'i}a, Facultad de Ingenier{\'i}a, Universidad Diego Portales, Av. Ej{\'e}rcito 441, Santiago, Chile}

\begin{abstract}
\label{Abstract}

ALMA measurements for 93 \textit{Herschel}-selected galaxies at $1.1 \leqslant z \leqslant 1.7$  in COSMOS reveal a sizable ($>29$\%) population with compact star formation (SF) sizes, lying {on average $> \times 3.6$ below} the optical stellar mass ($M_{\star}$)-size relation of disks.
This sample widely spans the star-forming Main Sequence (MS), having $10^{8} \leqslant M_{\star} \leqslant 10^{11.5} \ M_{\odot}$ and $20 \leqslant SFR \leqslant 680 \ M_{\odot} \rm yr^{-1}$. 
The 32 size measurements and 61 upper limits are measured on ALMA images that combine observations of CO(5-4), CO(4-3), CO(2-1) and $\lambda_{\rm obs} \sim 1.1-1.3 \ \rm mm$ continuum, all tracing the star-forming molecular gas. 
These compact galaxies have instead normally extended $K_{band}$ sizes, suggesting strong specific $SFR$ gradients.
Compact galaxies comprise the $50\pm18 \%$ of MS galaxies at $M_{\star} > 10^{11} M_{\odot}$. 
This is not expected in standard bi-modal scenarios where MS galaxies are mostly steadily-growing extended disks. 
We suggest that compact MS objects are early post-starburst galaxies in which the merger-driven boost of SF has subsided. They retain their compact SF size until either further gas accretion restores pre-merger galaxy-wide SF, or until becoming quenched. 
The fraction of merger-affected SF inside the MS seems thus larger than anticipated and might reach $\sim 50$\% at the highest $M_{\star}$. 
The presence of large galaxies above the MS demonstrates an overall poor correlation between galaxy SF size and specific $SFR$. 
\end{abstract}


\section{Introduction} \label{sec:intro}
The stellar mass ($M_{\star}$)-size relation provides important insights on $M_{\star}$ assembly processes.
Observational studies in the rest-frame optical/near-IR have shown that the dependence of size on $M_{\star}$ varies with galaxy type: late-type, star-forming disks (LTGs) have sizes which are mildly dependent on $M_{\star}$ and are larger than quiescent early-type galaxies \citep[ETGs, ][]{vanDerWel14}. Instead, local (U)LIRGs and high-redshift starburst galaxies (SBs) have very compact gas/star formation rate ($SFR$) sizes, driven by the ongoing merger \citep{Tacconi08}. Thus, sizes appear to be related to star formation activity. 

To better understand how galaxies are growing, optical/near-IR continuum studies, tracing $M_{\star}$, ought to be complemented by structural analyses of the star-forming component.
Observations of the H$\alpha$ emission show that low-to-moderately obscured star formation (SF) at $z \gtrsim 1$ takes place generally within disks similarly or even more extended than $M_{\star}$ \citep[e.g.][]{Nelson16b} although dust attenuation might reduce their effective radius \citep{Nelson16a}.

A different picture is emerging from studies imaging high-$z$ galaxies at long wavelengths. 
Various works show that obscured SF is hosted in compact regions of massive Main Sequence (MS) galaxies \citep{Tadaki17, Elbaz18}, sub-millimeter galaxies \citep[SMGs,][]{Miettinen17, Hodge16, Fujimoto18} and ``blue nuggets'' \citep[][]{Barro16}. 
On the other hand, stacking ALMA images in the $uv$ plane reveals that ``typical'' high-$z$ galaxies 
host extended obscured SF \citep{Lindroos16, Zanella18}. 

In this Letter we constrain the ALMA sizes for a statistical sample of 93 IR-selected galaxies at $1.1 \leqslant z \leqslant 1.7$ in COSMOS spanning a wide range in $M_{\star}$ and $SFR$.
We use a \cite{Chabrier} IMF and a standard cosmology ($H_{\rm 0} = 70 km s^{-1} Mpc^{-1}$, $\Omega_{\rm m} = 0.3$, $\Omega_{\rm \Lambda} = 0.7$).

\section{Sample and measurements} \label{sec:ALMA}

\subsection{ALMA observations and size measurements}
\label{sub_sec:Size measurements}
We use ALMA observations targeting the CO(5-4) transition and $\lambda_{\rm obs} \sim 1.3 \ \rm mm$ underlying continuum (circularized beam $\sim 0.7''$) in 123 far-IR selected galaxies with $1.1 \le z_{\rm spec} \le 1.7$ in COSMOS (Program-ID 2015.1.00260.S, PI Daddi). Galaxies were selected requiring a $3\sigma$ detection at 100 and/or 160 $\mu$m in PACS-\textit{Herschel} catalogs from the PEP survey \citep{Lutz}, implying $L_{\rm IR} \gtrsim 10^{12} \ L_{\odot}$.
We also have CO(2-1) observations for a subset of 75 galaxies ($\sim 1.5''$ beam; Program-ID 2016.1.00171.S, PI Daddi) and CO(4-3), [CI] and underlying continuum observations for 29 objects \citep[$\sim 2 ''$ beam, ][]{Valentino18}.
We analyze these tracers together to increase the size measurements accuracy. Here we focus on the 93 sources detected at $\gtrsim 5\sigma$ in at least one continuum or line tracer, for which we can estimate reliable sizes or robust upper limits. For a full description of the data-set, data reduction and measurements, we refer to future papers. 

To analyze the data we use GILDAS-based\footnote{http://www.iram.fr/IRAMFR/GILDAS} scripts.
The scripts optimize spectral extraction spatial position (constant over all data-sets) and emission line ranges iteratively for each detected emission, based on the recovered signal (see \citealt{Daddi15} and \citealt{Valentino18}).

\begin{figure*}
    \centering
\subfloat{\includegraphics[width=0.5\textwidth]{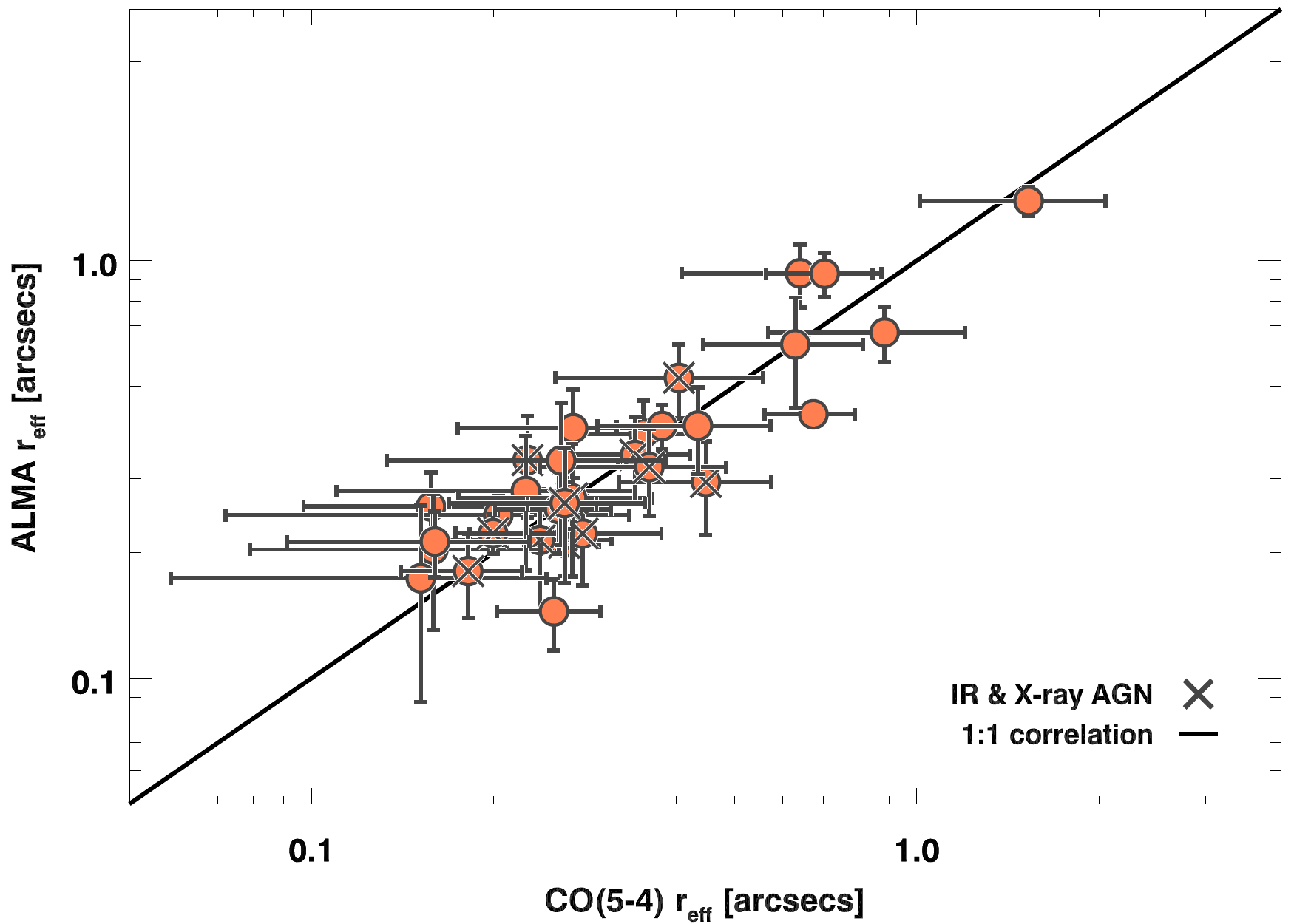}}\hfill
\subfloat{\includegraphics[width=0.5\textwidth]{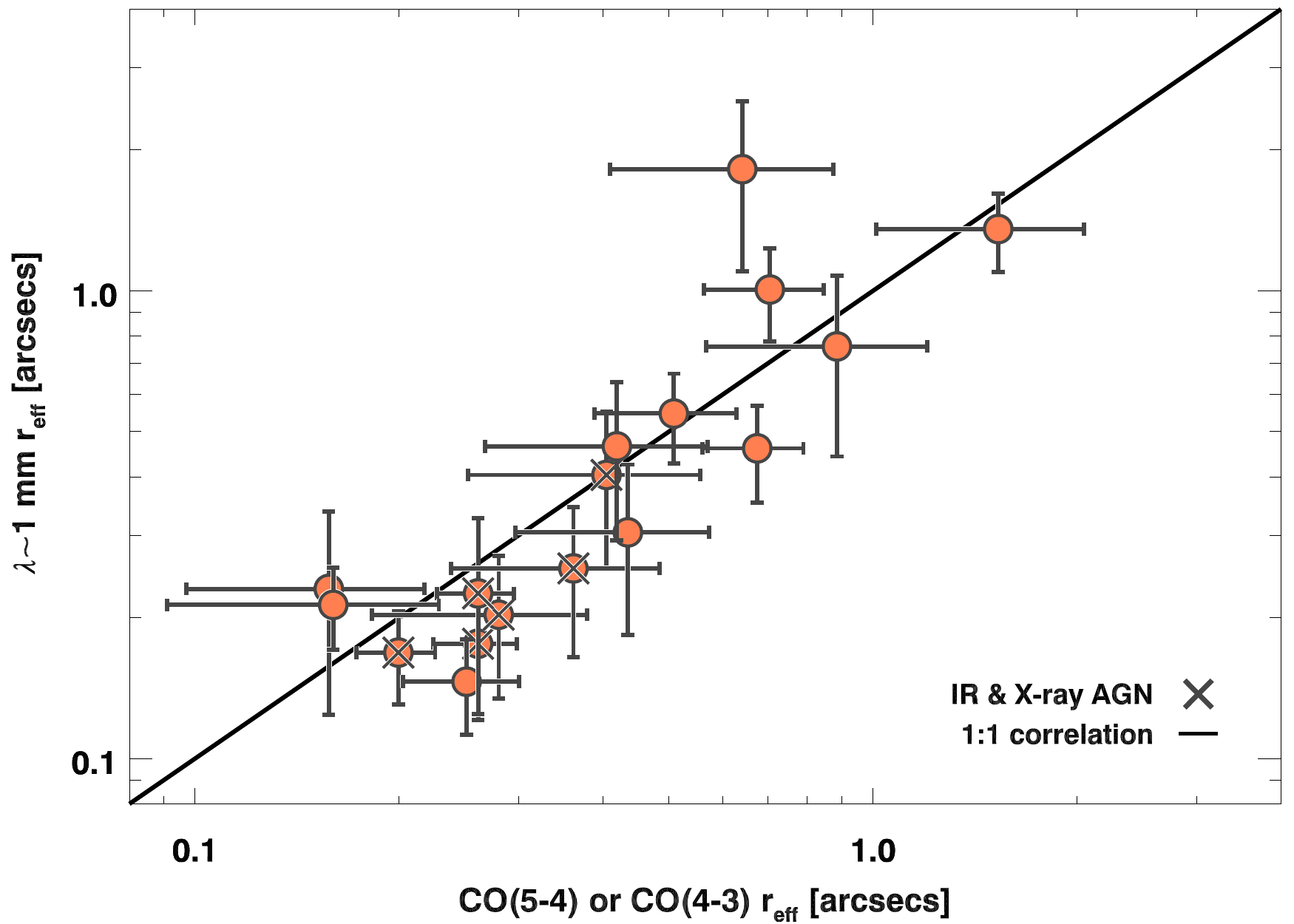}}\hfill
\subfloat{\includegraphics[width=0.7\textwidth]{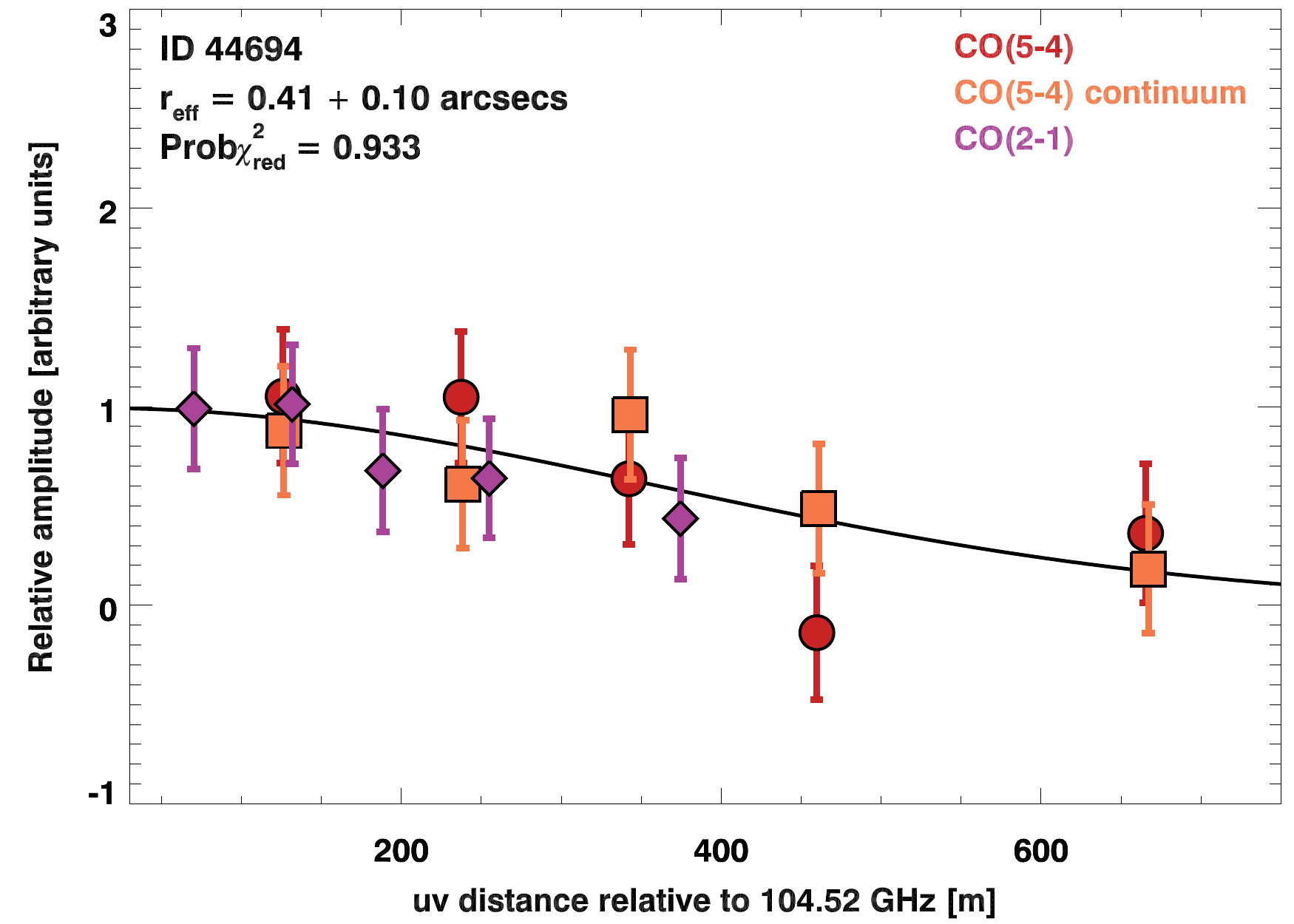}}
\caption{Characterization of our size measurements. 
\textit{Upper-left:} CO(5-4) versus ALMA sizes. \textit{Upper-right:} high-J CO versus dust continuum sizes.
\textit{Bottom:} Amplitude as a function of the \textit{uv} distance for a galaxy in the sample. The black line is the best-fit Gaussian profile.}
\label{fig0:Optimal_COSize}
\end{figure*}

To measure sizes, the scripts extract the signal amplitude as a function of the \textit{uv} distance from each tracer, combining their signals scaling the \textit{uv}-distances to a common frequency and { marginalizing over a} free normalization constant for each tracer.
The galaxy best-fit size {(defined as the effective radius $r_{\rm eff} = FWHM/2$)} and its $1 \sigma$ uncertainty is determined by comparing the \textit{uv} distance vs. amplitude distribution to circular Gaussian models (see Fig. \ref{fig0:Optimal_COSize}). 
The goodness of the fit is estimated from the $\chi^2$ minimization as a function of the size.
Unresolved sources would show a constant amplitude profile in the \textit{uv}-plane.
To quantify the probability of each galaxy to be unresolved ($P_{\rm unres}$) the scripts compare the best-fit $\chi^2$ to the $\chi^2$ for a point source. A source is considered to be resolved when $P_{\rm unres} \leqslant 10 \%$. 
We obtain 32 resolved sources (average size/size-error $\sim 5.3$) with this threshold, which is determined \textit{a-posteriori} in a way that we statistically expect only 0.5 of these to be spuriously resolved, on average. 
For 61 sources we cannot derive robust size constraints, and we show these as 1$\sigma$ upper limits in our plots.
Depending on the combined SNR of the emission line/continuum detection, size upper limits vary from being quite stringent ($r_{\rm eff} \sim 0.10''$) to fairly loose. The median upper limit is $r_{\rm eff} = 0.28 ''$. 
30 sources have no significant detection in the ALMA data-set.
We use Monte Carlo simulations to test this method. 
We create 1000 mock realizations of our data-sets by perturbing the best-fit model within the measured uncertainties in the \textit{uv}-amplitudes plot, assuming Gaussian noise and we measure sizes of each synthetic data-set. 
We simulate sources covering a broad range of sizes, from much smaller to comparable to the beam (similarly to what found for real galaxies) and over a range of SNR ratios as spanned by real galaxies. We find no significant systematics to within $5 \%$, much lower than any measurement errors.

Our procedure combines information from tracers with potentially different spatial distributions, requiring an intrinsic surface brightness distribution consistently extended within the uncertainties. 
This is confirmed with our data-set as sizes from independent tracers are in good agreement (see Fig. \ref{fig0:Optimal_COSize}).
We do expect consistency among individual sizes as all the tracers considered here are directly sensitive to the star-forming gas. The dust continuum and low-J CO emission are expected to be highly equivalent tracers of the gas \citep[e.g.][]{Magdis12}. 
The higher-J CO emission are sensitive to higher density gas, being closer proxies to the $SFR$ \citep[e.g.][]{Daddi15}. However, given the overall tight correlations between gas and $SFR$ \citep[e.g.][]{Sargent14}, we expect any intrinsic galaxy-wide size difference to be subtle.
Furthermore, Fig. \ref{fig0:Optimal_COSize} shows that the ALMA size is mostly driven by CO(5-4), typically detected with highest SNR and having also the smallest beam. 
Considering the CO(5-4) size measurements only does not affect our conclusions.

\subsection{$M_{\star}$, SFRs and $K_{\rm s}$-band sizes}

For the star-forming galaxies in our sample, we use $M_{\star}$ from \cite{Laigle16}.
Roughly $40 \%$ of galaxies are identified as AGN, as suggested by their X-ray emission and/or the torus contribution to the mid-IR part of the spectral energy distribution at $\gtrsim 5 \sigma$ significance.
For these AGN-hosts, we derive $M_{\star}$ using the code CIGALE \citep{Noll09} accounting for the AGN contribution as described in \cite{Circosta18}.
A subset of sources split-up in the $K_{\rm s}$-band image.
We compute $K_{\rm s}$-band magnitudes for each component of the pair using \textsc{galfit} as described below. We derive $M_{\star}$ for each component by re-scaling the total $M_{\star}$ to their $K_{\rm s}$-band magnitudes ratio. We then consider the galaxy closest to the ALMA position.

We derive $L_{\rm IR}$ over the range $\lambda \in [8 - 1000] \ \mu$m from the super-deblended catalog of \citealt[][]{Jin18} \citep[see also][]{Liu18}, including notably PACS and SPIRE \textit{Herschel} observations. Measurements of $L_{\rm IR}$ are based on the combination of the \cite{Magdis12} simplification of \cite{DraineLi07} dust models, and AGN models from \cite{Mullaney11}, and agree well with  $L_{\rm IR}$ values used when selecting the sample.
We estimate $SFR_{\rm FIR}$ using the \cite{Kenni98} conversion rescaled to a \cite{Chabrier} IMF. 
Four sources are pairs in CO(5-4). For these objects, we compute individual $SFRs$ by re-scaling the total $SFR_{\rm FIR}$ to the CO(5-4) fluxes ratio which correlates linearly with $L_{\rm IR}$  \citep{Liu15, Daddi15}.

We derive sizes in the near-IR rest-frame by fitting circular Gaussian profiles with \textsc{galfit} \citep{Peng10_GALFIT} on UltraVISTA $K_{\rm s}$-band images \citep{McCracken12}. 
The average seeing of these images ($\sim 0.7''$) is comparable to our CO(5-4) ALMA beam.
As for the ALMA measurements, $K_{\rm s}$-band sizes are the $r_{\rm eff}$ of the circular Gaussian profile\footnote{Using Sersic index $n=1$ does not affect the results.}. 
Nine galaxies in the sample are type-1 AGN, the $K_{\rm s}$-band size of the host is undetermined. 

\section{Results}\label{sec:Results}
\begin{figure*}[ht!!!]
\begin{center}
\centering
\epsscale{1.1}
\figurenum{2}
\plotone{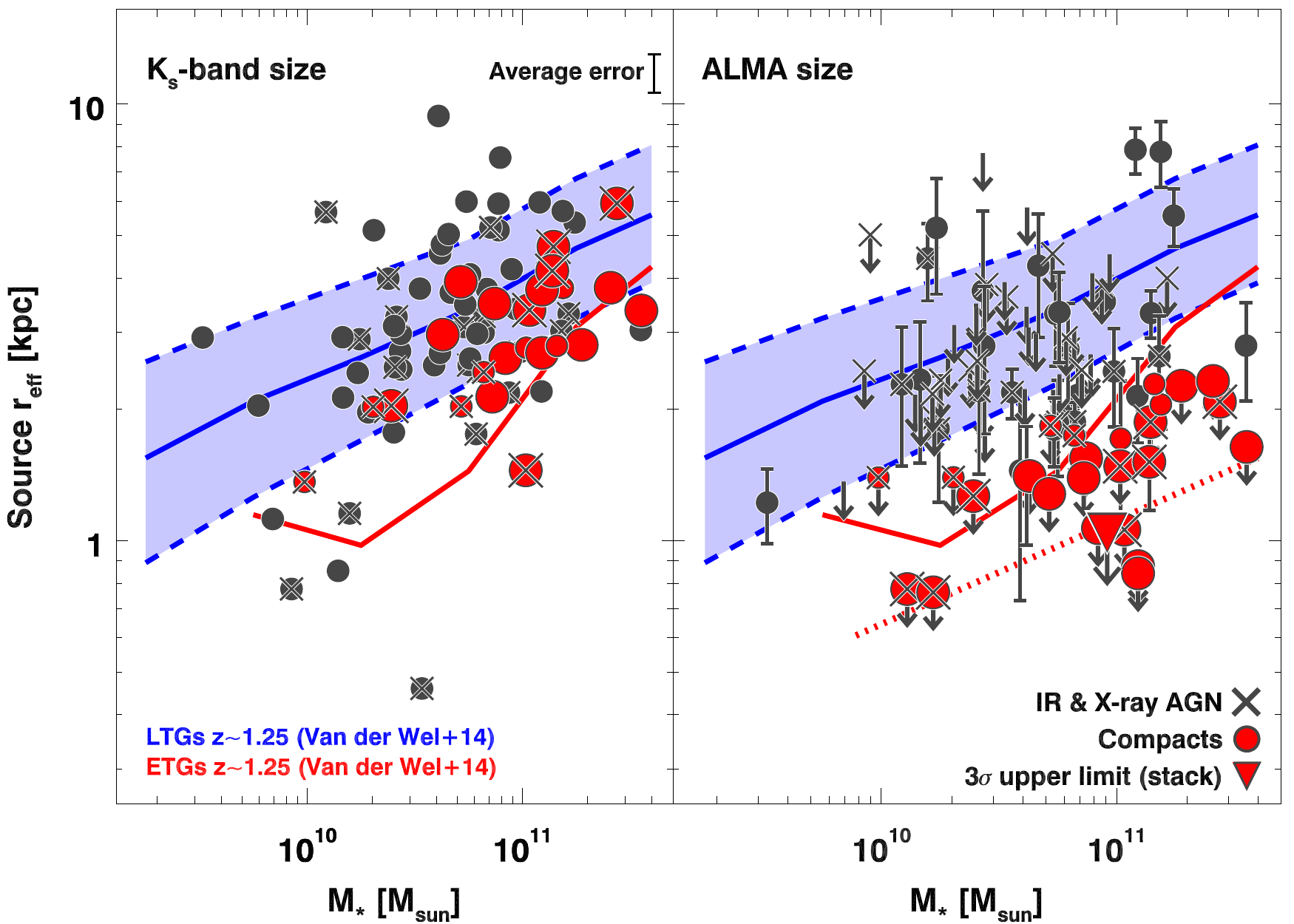}
\caption{ $M_{\star}$ versus $K_{\rm s}$-band (\textit{left}) and ALMA size (\textit{right}). 
The blue line and shaded area highlight the LTG $M_{\star}$-Size relation and its scatter at $z \sim 1.25$ from \cite{vanDerWel14}. The ETG $M_{\star}$-Size relation is shown for reference.
These consider circularized radii \citep[see Tab. 3 in][]{vanDerWel14} for consistency with our measurements.
Red circles mark ALMA compact galaxies in our sample. 
The downward red triangle highlights the 3$\sigma$ size upper limit for unresolved compact galaxies. }
\label{fig1:Mass_Size_Ks_ALMA}
\end{center}
\end{figure*}

We use $K_{\rm s}$-band and ALMA sizes to construct the $M_{\star}$-Size plane for the sample (see Fig. \ref{fig1:Mass_Size_Ks_ALMA}).
At these redshifts, the $K_{\rm s}$-band (rest-frame $\sim 1\mu$m) roughly traces the $M_{\star}$. Galaxies in our sample have the ``typical'' extension of star-forming disks in $K_{\rm s}$-band as they nearly all locate within the LTG relation of \cite{vanDerWel14}. Instead, ALMA measurements are skewed towards smaller sizes. 

We compute the $\chi^2$ of the data distribution as:
\begin{equation}
\chi^2 =  \sum_{i} \frac{(Size_{\rm ALMA, i} - Size_{\rm LTG, i})^2}{\sigma_{\rm ALMA, i}^2 + \sigma_{\rm LTG, i}^2}
\end{equation}
where $\sigma_{\rm LTG}$ and $\sigma_{\rm ALMA}$ are the LTG scatter and the $1 \sigma$ size error, respectively. We consider here all sources with their size measurements and errors (including those shown on plots as upper limits).
66 galaxies are consistent with the LTG relation whereas 27 are more compact than expected in an average implementation of the LTG. These 27 are our ``compact sample'' reported in red in the plots. Of these, 18 are inconsistent with the LTG relation at $\geqslant 3 \sigma$ significance and are highlighted with larger symbols. 
To verify the robustness of our results, we compute the compact fraction via a survival analysis approach through the non-parametric Kaplan-Meier \mbox{estimator}, using $1\sigma$ upper limits. The lower 1$\sigma$ boundary in the compact fraction derived in this way is in good agreement with our conservative estimate described above. 
{To obtain an improved average constrain on unresolved compact galaxies, we combine their observations in the \textit{uv} distance vs. amplitude plane measuring the size as in Sect. \ref{sub_sec:Size measurements}. This results in an unresolved average source with a 3$\sigma$ upper limit $r_{\rm eff} = 0.13 ''$ (downward triangle in Fig. \ref{fig1:Mass_Size_Ks_ALMA} ).}

\begin{figure*}[ht!]
\begin{center}
\centering
\epsscale{1.1}
\figurenum{3}
\plotone{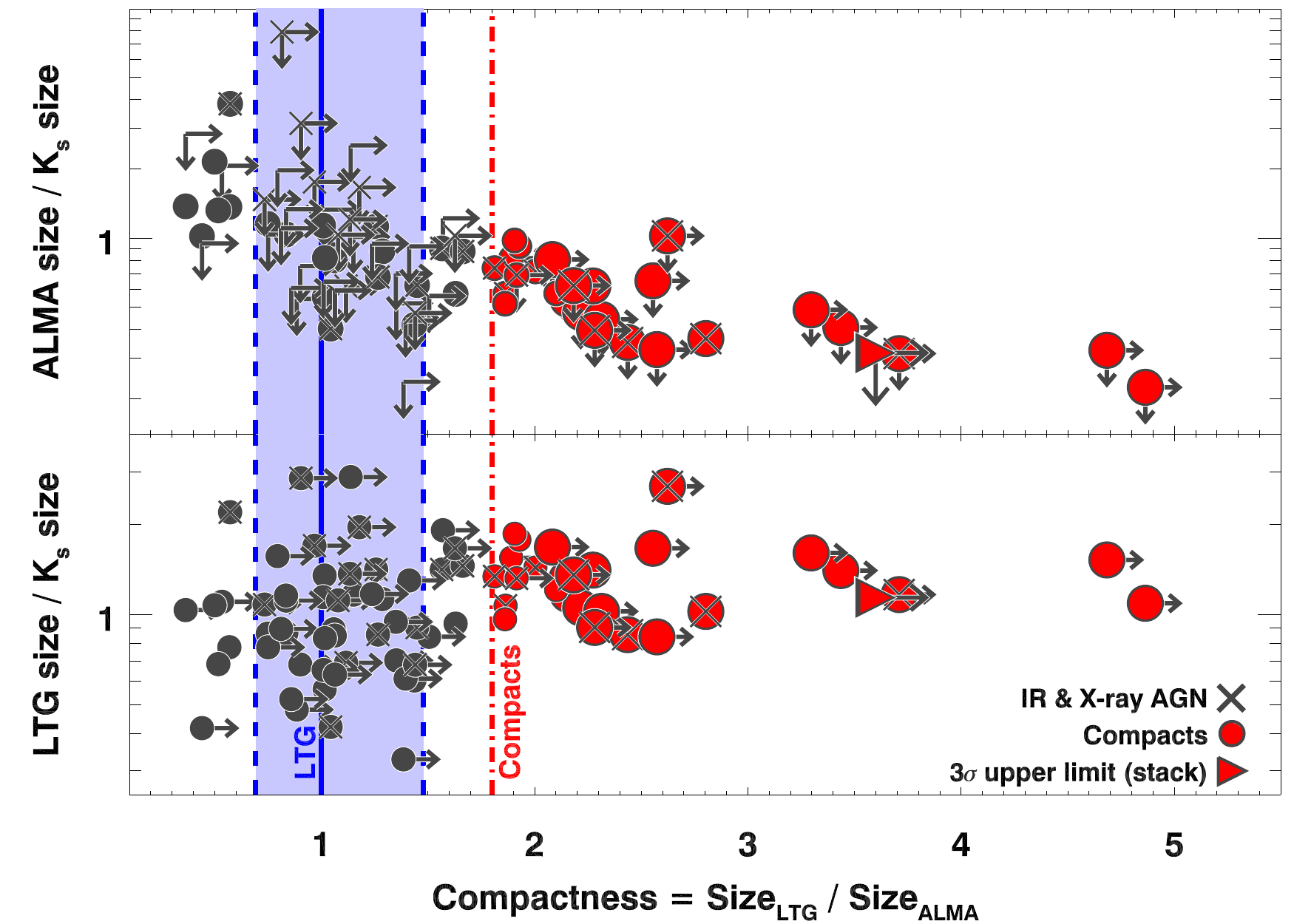}
\caption{\textit{Upper panel:} ALMA versus $K_{\rm s}$-band size ratio as a function of the compactness.
\textit{Lower panel:} LTG versus $K_{\rm s}$-band size ratio as a function of the compactness.
The red dash-dotted line marks the threshold below which we classify galaxies as compacts.
The LTG relation at the average $M_{\star}$ of the sample is shown for reference  
\label{fig2:CO_vs_Ks_vs_LTGdistance}}
\end{center}
\end{figure*}

{To compare the $M_{\star}$ and molecular gas size on an individual basis, we plot the $K_{\rm s}$-band-to-ALMA and the LTG-to-$K_{\rm s}$-band size ratios as a function of the compactness (Fig. \ref{fig2:CO_vs_Ks_vs_LTGdistance}). 
The compactness is defined as the ratio between the LTG size at the galaxy $M_{\star}$ and the ALMA size. 
Even if compactness measurements are lower limits for most compact galaxies, the y-axes in this figure show that compact galaxies have a ALMA component smaller than the $K_{\rm s}$-band. 
By considering the average constraints, the compact population is $> 3.2 \times$ smaller in ALMA than in the $K_{\rm s}$-band at $3 \sigma$.
The lower panel shows that most galaxies have LTG-to-$K_{\rm s}$-band size ratio around one within a factor of two.}

\section{Discussion}\label{sec:Discussion}

The ALMA size is measured on images combining star-forming gas tracers and tightly correlates with the 
\mbox{CO(5-4)} emission (see Fig. \ref{fig0:Optimal_COSize}) being sensitive to the star-forming molecular gas \citep{Daddi15, Liu15}.
This implies that compact galaxies in our sample have most of the SF activity occurring in a nuclear, compact region\footnote{The observed UV emission is a negligible fraction of $SFR_{\rm FIR}$ in all our sources .}. 
In the following, we will consider this feature as arising from mergers and interactions \citep{BarnesHernquist92}. 
Alternative mechanisms such as violent disk instabilities \citep{Bournaud16} cannot dramatically reduce the global SF size of galaxies to factors of several below that of normal disks. 
Anomalous streams of gas may also drive large gas concentrations in the nucleus \citep{Dekel14}. However, the occurrence of those events has never been proven observationally. Moreover, mergers or fly-by are often required to trigger those streams in simulations making this scenario effectively coincident with our proposed interpretation.

\begin{figure*}
\figurenum{4}
\centering
\subfloat{\includegraphics[width=0.5\textwidth]{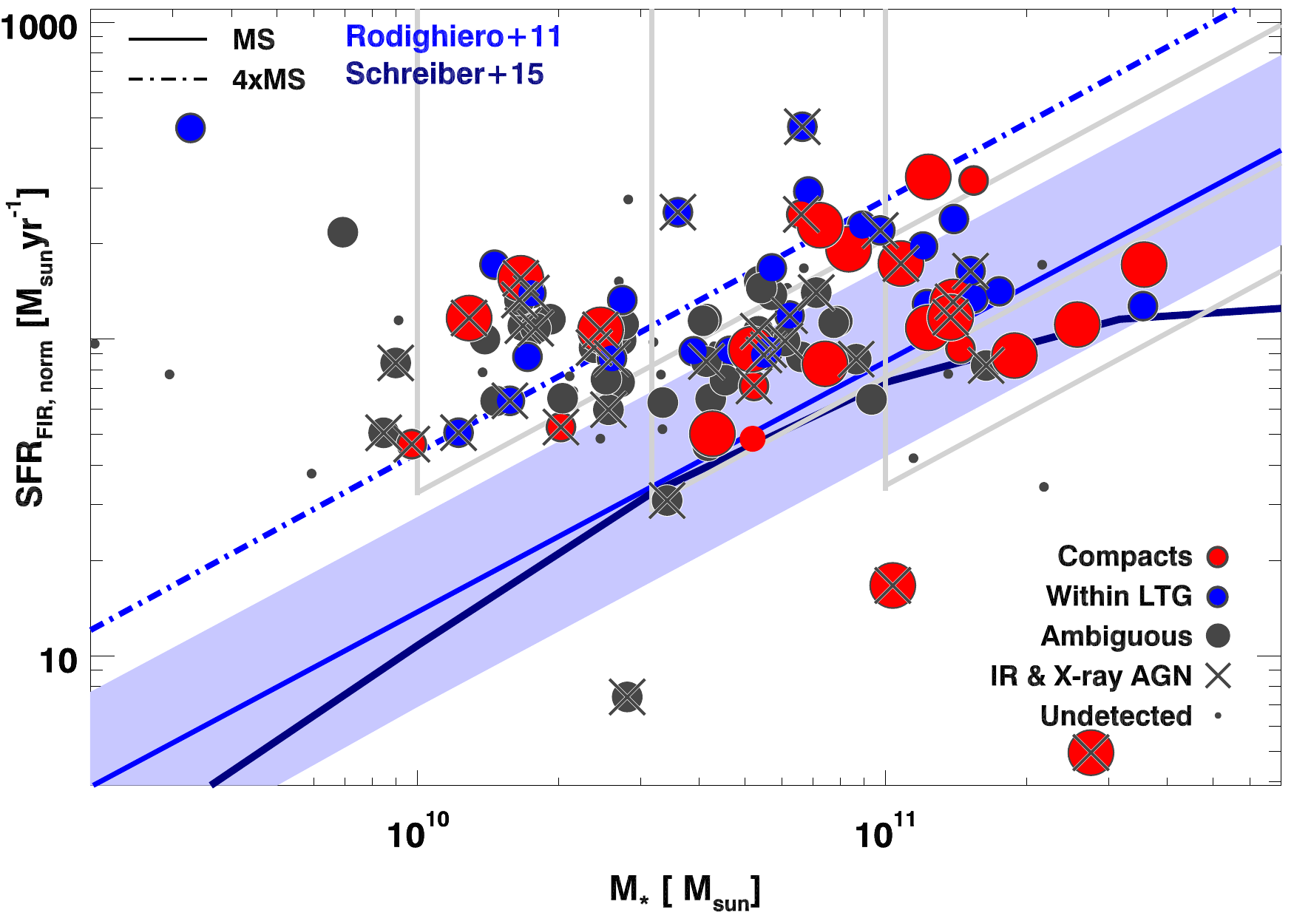}}\hfill
\subfloat{\includegraphics[width=0.5\textwidth]{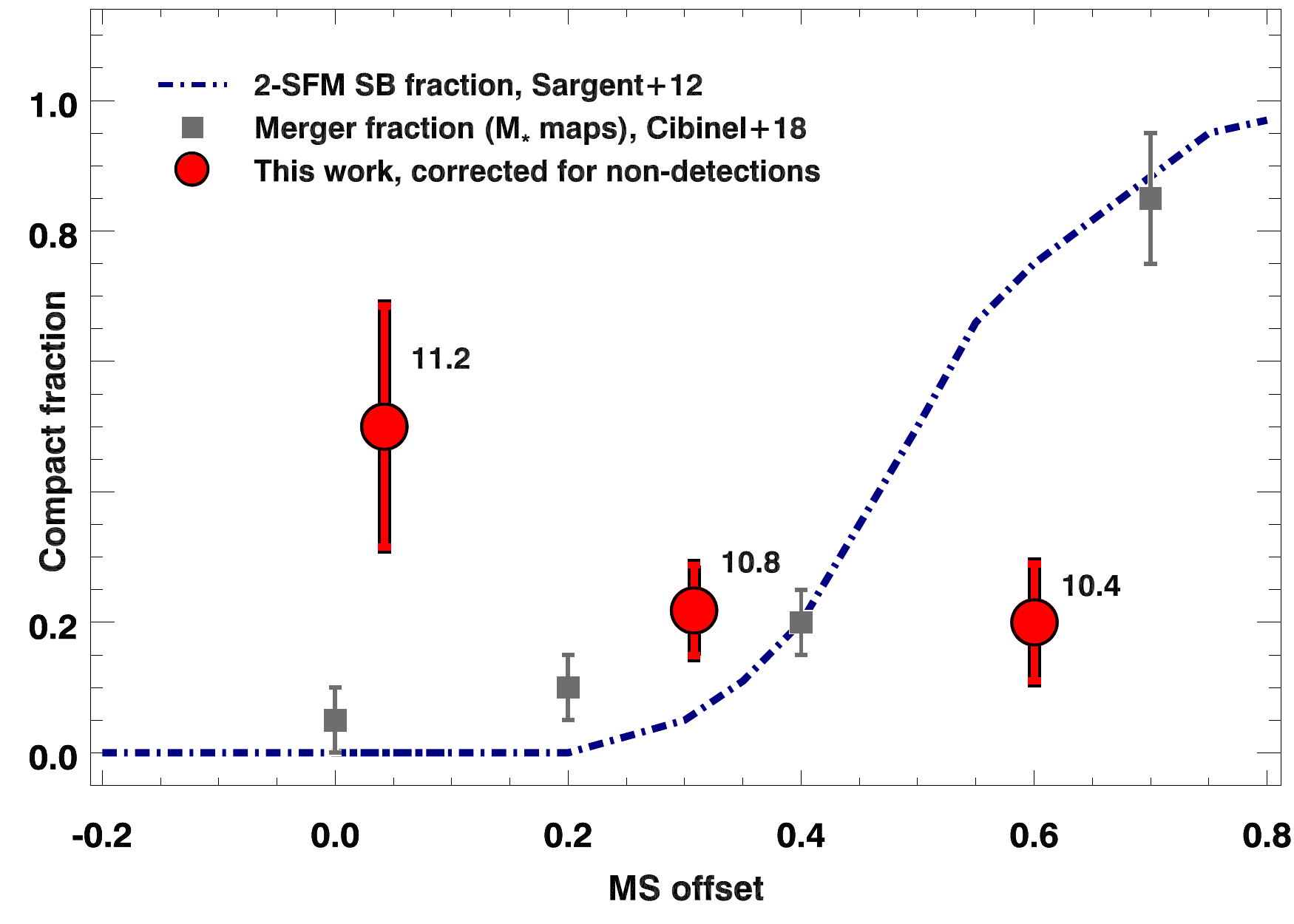}}\hfill
\caption{\textit{Left:} $SFR$ as a function of $M_{\star}$ for the 123 galaxies from our main ALMA program. Blue circles highlight galaxies with size measurements within the LTG $M_{\star}$-Size relation.
Red circles indicate the compacts and bigger circles are below the LTG relation at $3 \sigma$.
Grey large circles are upper limits within the LTG relation, for which we cannot place constraints on the compactness.  
\textit{Right:} Fraction of compact galaxies as a function of the MS distance. Numbers indicate the average $M_{\star}$ of each bin.}
\label{fig3:MS_Compact_fraction_MS}
\end{figure*}

Mergers are expected to enhance the galaxy $SFR$, as suggested in the local Universe \citep{SandersMirabel} and at high-redshift \citep{Rodighiero11}. 
As such, we might expect to find compact galaxies preferentially above the MS.
In Figure \ref{fig3:MS_Compact_fraction_MS}-left we plot the $SFR_{\rm FIR}$ as a function of $M_{\star}$ for our sample, including information on the molecular gas size.  
Surprisingly, we find no clear correlation between the molecular gas size and the MS position.
This is in contrast with the regularity seen from H$\alpha$ \citep[e.g.][]{Nelson16b} and possibly in the dust-obscured SF component. 
For example, \cite{Rujopakarn16} show that $SFR$ and $M_{\star}$ are similarly extended in dusty MS galaxies, supporting the existence of wide-spread obscured star formation within the MS. 
Similarly, \cite{Miettinen17} show that SMGs within the MS are extended and disk-like.

In our study, several MS-outliers are consistent with the LTG relation, at odds with the idea that off-MS galaxies are compact merger-driven starbursts \citep{Silverman15,  Silverman18_Pacs787, Silverman18,Calabro18}. 
This is consistent with the idea that these galaxies are gas-rich objects with disk-wide starbursts  \citep[e.g.]{Scoville16}.
We suggest however that at least some of these objects might be merging pairs in a pre-coalescence phase, individually unresolved with our $\sim 0.7'' - 1.5 ''$ beam, as HST imaging supports in some cases. This would be consistent with results from hydrodinamical simulations \citep{Perret14} and radio sizes of pre-coalescence MS-outliers \citep{Calabro19}. 
This is also suggested by the case study of a MS-outlier with a $\sim 10 \ kpc$ CO disk at $\sim 2''$ resolution splitting into two $\sim 1 \ kpc$ disks at higher resolution \citep{Silverman18_Pacs787}.
Follow-up at higher spatial resolution will allow us to distinguish these scenarios.

More puzzling is the presence of compact galaxies within the MS.
A quantification of the prevalence of these sources is essential, as, e.g., the bi-modal population model of \cite{Sargent12} predicts the existence of some SBs within the MS.
We thus compute the fraction of compact galaxies in three $sSFR$ bins on the MS (light grey lines in Fig. \ref{fig3:MS_Compact_fraction_MS}-left) and we compare this quantity with expectation from the \cite{Sargent12} model. 
We plot the fractions of compact galaxies as a function of the MS position in the right panel of Fig. \ref{fig3:MS_Compact_fraction_MS}.
Compact galaxies are the $\sim 50$\% of the MS population at $M_{\star} > 10^{11} M_{\odot}$, with a negligible uncertainty in this bin coming from size measurement uncertainties (but still limited by statistics). 
As we consider compact galaxies to be most likely mergers, this result is not consistent with the canonical MS paradigm whereby MS galaxies are disks experiencing steady-state growth.

We note that we observed the $\sim 1 \%$ of all COSMOS galaxies above $>10^{12}L\odot$, notably requiring a spectroscopic redshift to enable the CO follow-up.  
However, the biasing effect of the spectroscopic sample on sizes should be small. 
Nevertheless, we account for selection effects that might bias our observations against the most extended galaxies, which would result systematically in lower SNR for a fixed luminosity. Conservatively, we consider undetected sources in our sample (grey dots in Fig \ref{fig3:MS_Compact_fraction_MS}-left) as consistent with the LTG relation, together with size upper limits within the LTG relation. 
We emphasize that ALMA size measurements include low-J CO observations and dust continuum tracing lower density, possibly more extended, molecular gas. Although significant size variations across tracers are not detected in our data-set (see Sect. \ref{sec:ALMA}), the presence of such a bias would further strengthen our conclusions. We explore in the following possible causes for the excess of compact galaxies within the MS. 

A possibility is that we are observing galaxies in which the merger has not been capable of triggering intense SB activity due to the enhanced gas fractions, as suggested by simulations \citep{Fensch17}.
While this might happen in practice, these mergers with failed bursts would likely be classified as such by morphological methods, but our compact fraction significantly exceeds the merger fraction measured on $M_{\star}$ maps \citep[][see also Fig. \ref{fig3:MS_Compact_fraction_MS}]{Cibinel18}. 
Also, we expect that mergers not triggering strong starbursts would also fail at producing very compact SF cores \citep{Bournaud15}.
\begin{figure}[ht!]
\begin{center}
\centering
\epsscale{1.2}
\figurenum{5}
\plotone{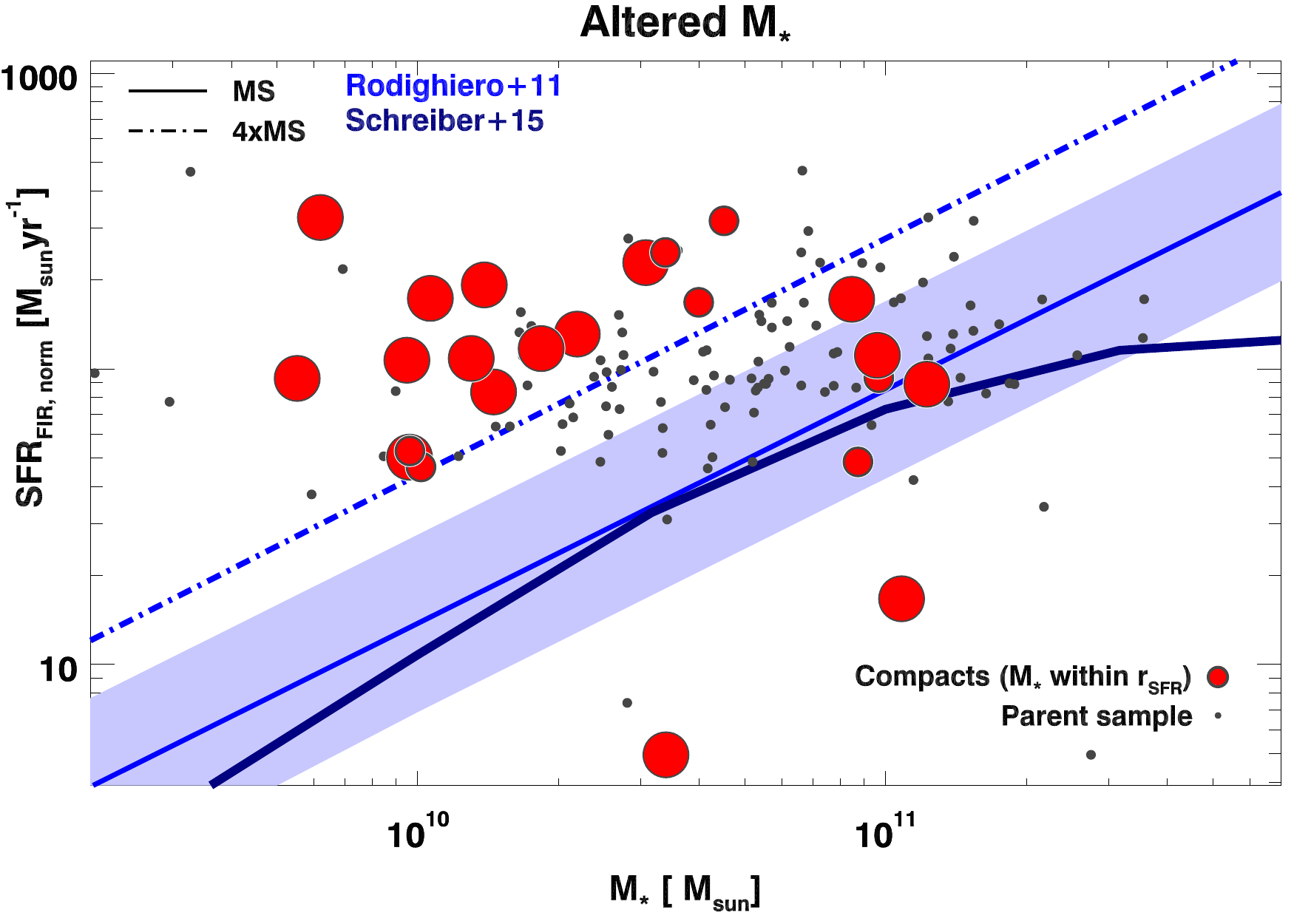}
\caption{Same as Fig.\ref{fig3:MS_Compact_fraction_MS}-left but with $M_{\star}$ of compact galaxies re-scaled to the obscured $SFR$ extension as traced by ALMA.
\label{fig4:MS_Mstar_rescaled}}
\end{center}
\end{figure}

We notice though that the different $M_{\star}$ and $SFR$ extensions in compact galaxies (see Fig. \ref{fig2:CO_vs_Ks_vs_LTGdistance}) implies strong specific $SFR$ ($sSFR$) gradients. Accounting for only the $M_{\star}$ within the star-forming cores by scaling the total $M_{\star}$ by $r^{2}_{K_{\rm s}}/r^{2}_{\rm ALMA}$, compact galaxies actually host starbursts in their nuclei, with $sSFRs$ that would place them off the MS (see Fig. \ref{fig4:MS_Mstar_rescaled}). These nuclear starbursts are not intense enough for the whole galaxy to classify it as MS outlier. We thus suggest compact MS galaxies to be post-starburst systems in which the burst has subsided their peak levels. 
We are likely sampling different phases of the merger.  In fact, the typical duration of merger-triggered bursts is $\sim$50-100 Myr \citep[e.g.]{DiMatteo08}, and as soon as the activity diminishes after consuming a large fraction of the gas reservoir, the $SFR$ is likely to remain centrally peaked for longer times. 
The typical timescales for gas re-accretion are in fact much longer \citep[$\sim 0.5-1 \ Gyr$) at these redshifts][]{Sargent12} and similar to galaxies doubling time \citep{Lilly13}. It is also possible that, following the merger, these galaxies will evolve to become quiescent. 
Hence, we would expect these post-SB galaxies to be more numerous than SBs, and located inside the MS. 
Galaxies in this post-SB phase would not be selectable by the \cite{Cibinel18} method, which is sensitive to mergers up to the coalescence, i.e. until the system retains $M_{\star}$ maps asymmetries \citep[see also][]{Cibinel15}.  

All in all, we find no clear correlation between the molecular gas size and the MS position and $\geqslant 50$\% of MS galaxies above $M_{\star} \sim 10^{11} \ M_{\odot}$ are compact in the molecular gas. 
Such high percentage of compact galaxies within the MS is in line with previous literature results, showing that up to the $\sim 60\%$ of MS galaxies above $10^{11} M_{\odot}$ host compact molecular gas cores \citep{Tadaki17, Elbaz18}.
We suggest that compact galaxies at the massive end of the MS might represent post-SB mergers which have ended the starburst phase in the off-MS region but are still mainly forming stars in a compact nucleus or a post-merger compact disk \citep[see e.g. Fig. 2 in][]{Fensch17}.
This suggests that the contribution of bursty (merger-driven) SF to the cosmic $SFR$ density is larger than previously anticipated \citep[$\sim 15 - 20$\%, e.g.,][]{Rodighiero11},  and might reach up to $\sim50$\% at the high-mass end where SF disks become rarer (as also expected because we are sampling beyond the characteristic $M_{\star}$ of their mass functions).
The ALMA compactness might provide a tool to select MS galaxies in an early post-SB phase, complementary to absorption lines or color selection criteria selecting objects already below the MS \citep{Wild09}.
Further studies on the stellar population properties of the compact population will allow us to confirm this hypothesis. If confirmed, this ``early post-SB'' population might open scenarios for understanding the role of merger-driven starbursts in the galaxy life-cycle and may possibly provide insights on the passivization mechanisms at high-redshift. 

\acknowledgements{\small{We acknowledge funding INAF PRIN-SKA 2017  1.05.01.88.04,  the Chinese Academy of Sciences President's International Fellowship Initiative 2018VMA0014, 
  the Villum Fonden  grant 13160, the Cosmic Dawn Center of Excellence by the Danish National Research Foundation,   ERC Consolidator  648179,
 National Science Foundation AST-1614213,
the Spanish Ministry of Economy and Competitiveness 2014 Ramon y Cajal  MINECO RYC-2014-15686,
the National Key R\&D Program of China (2017YFA0402704), the CAS Key Research Program of Frontier Sciences and NSFC 11420101002.}}

\end{document}